\magnification=\magstep1
\centerline{\bf Concerning the Differences Between Zero Mass Dirac 
and Majorana Neutrinos}
\medskip
\centerline{Y. N. Srivastava and A. Widom}
\centerline{Physics Department, Northeastern University, Boston MA 02115}
\centerline{and}
\centerline{Physics Department \& INFN, University of Perugia, Perugia, Italy}
\medskip
\centerline{\it ABSTRACT}
\medskip
That a Majorana neutrino theory is different from a Dirac 
neutrino theory (even in the zero mass limit) is proved in 
two sentences.
\medskip
\centerline{\bf Fermion Statistics}
\medskip
For the reactions (Dirac) $e^+e^-\to \nu \bar{\nu}$ and (Majorana)  
$e^+e^-\to \nu \nu$ wherein the two final particles are presumed 
to have opposite helicity, in the Dirac case the neutrino 
and anti-neutrino are {\it different particles} so that   
$\Psi_{Dirac}({\bf p}_1,\lambda_1 ;\bar{{\bf p}}_2,\bar{\lambda }_2)$
(in momentum-helicity) has no particular particle anti-particle 
exchange symmetry, while in the Majorana case the two neutrinos 
are identical fermion particles with an exchange 
anti-symmetric wave function {\it in the 
center of mass frame}[1]
$$
\Psi_{Majorana } 
({\bf p}_1={\bf p},\lambda_1;{\bf p}_2=-{\bf p},\lambda_2)=
-\Psi_{Majorana } 
({\bf p}_1=-{\bf p}, -\lambda_2;{\bf p}_2={\bf p},-\lambda_1).  \eqno(1)
$$ 
Thus, for the Dirac case with a fixed angle of scattering, 
the probability of having a left helicity neutrino can and does differ 
from the probability for having a right helicity anti-neutrino 
$$
\Big({d\sigma \over d\Omega}\Big)_{Dirac\ \nu \ Left}
\ne \Big({d\sigma \over d\Omega}\Big)_{Dirac\ \bar{\nu}\ Right}
\ \ \ (in\ general) \eqno(2)
$$ 
and for the Majorana case with a fixed angle of scattering,  
the probability of having a left helicity neutrino is identical 
to the probability of having a right helicity neutrino 
$$
\Big({d\sigma \over d\Omega}\Big)_{Majorana\ \nu\ Left}=
\Big({d\sigma \over d\Omega}\Big)_{Majorana\ \nu\ Right}. \eqno(3)
$$

This non-perturbative difference (transcending particular 
parity violating Lagrangian models) between the Dirac and Majorana 
cases is true even in the zero mass limit. We disclaim 
responsibility for work purporting to describe more than one 
Majorana fermion without employing a properly anti-symmetric 
neutrino wave function. Finally, we have greatly benefited by 
reading recent debates[2-6].
\medskip

\centerline{\bf REFERENCES}
\medskip
\par \noindent
1. V. B. Berestetskii, E. M. Lifshitz and L. P. Pitaevski, 
{\it Landau and Lifshitz Course of Theoretical Physics, Vol. 4, 
2nd Edition, Quantum Electrodynamics}, Chapt. VII, Sec.69, 
Pergamon Press, Oxford (1982). 
\par \noindent
2. R. Plaga, hep-ph/9610545 (1996).
\par \noindent
3. S. Hannestad, hep-ph/9701216 (1997)
\par \noindent
4. B. Kaiser, hep-ph/9703294 (1997)
\par \noindent
5. A. Hoefer, hep-ph/9705362 (1997).
\par \noindent
6. S. H. Hausen, hep-ph/9708359 (1997)
\par \noindent

\bye